\documentclass[conference]{IEEEtran}
\IEEEoverridecommandlockouts
\usepackage{cite}
\usepackage{tabularx}
\usepackage{amsmath,amssymb,amsfonts}
\usepackage{float} 
\usepackage{algorithmic}
\usepackage{graphicx}
\usepackage{placeins} 
\usepackage{textcomp}
\usepackage{booktabs}
\usepackage{caption}  
\usepackage{booktabs}
\usepackage{xcolor}
\usepackage[numbers]{natbib}
\def\BibTeX{{\rm B\kern-.05em{\sc i\kern-.025em b}\kern-.08em
T\kern-.1667em\lower.7ex\hbox{E}\kern-.125emX}}
\begin{document}
\author{
Zhaorui Sun\textsuperscript{1,2*}\thanks{\textsuperscript{*}Equal contribution}, 
Yihao Chen\textsuperscript{2,3*}, 
Jialong Wang\textsuperscript{2}, 
Minqiang Xu\textsuperscript{2}\textsuperscript{\dag}\thanks{\textsuperscript{\dag}Corresponding Author},
Jianbo Zhan\textsuperscript{2\dag},
Lei Fang\textsuperscript{2}, 
Sian Fang\textsuperscript{2}, 
Lin Liu\textsuperscript{2}\\[2.5ex]  
\textsuperscript{1}School of Computer Science and Technology, Xinjiang University, Urumqi, China \\
\textsuperscript{2}Hefei iFly Digital Technology Co. Ltd., Hefei, China \\
\textsuperscript{3}University of Science and Technology of China, Hefei, China \\[0ex] 
}

\title{Enhancing Self-Supervised Speaker Verification Using Similarity-Connected Graphs and GCN}

\maketitle

\begin{abstract}
With the continuous development of speech recognition technology, speaker verification (SV) has become an important method for identity authentication. Traditional SV methods rely on handcrafted feature extraction, while the introduction of deep learning has significantly improved system performance. However, the scarcity of labeled data still limits the widespread application of deep learning methods in SV. Self-supervised learning, by mining the latent information in massive unlabeled data, enhances the model's generalization ability and has become a key technology to address this issue.

DINO  is an efficient self-supervised learning method that generates pseudo-labels from unlabeled speech data through clustering algorithms, providing support for subsequent training. However, the clustering process may produce noisy pseudo-labels, which can reduce the overall recognition performance of the system and restrict further improvement of the model's performance.

To address this issue, this paper proposes an improved clustering framework based on similarity connection graphs and Graph Convolutional Networks (GCN). By leveraging GCN's strength in modeling structured data and incorporating the relational information between nodes in the similarity connection graph, the clustering process is optimized, improving the accuracy of pseudo-labels and thereby enhancing the robustness and performance of the self-supervised speaker verification system. Experimental results show that this method can significantly improve system performance and provide a new approach for self-supervised speaker verification.
\end{abstract}


\section{Introduction}
Speaker verification (SV) is a biometric technology that verifies an individual's identity by analyzing speech features. The core task of SV is to determine whether the input speech matches the voiceprint of a specific individual stored in a database, thereby enabling identity authentication or access control. As a key application of speech interaction technology, SV has been widely used in security systems, voice assistants, financial transactions, and smart homes, and its development directly impacts both system security and the convenience of user experience.

Traditional SV systems primarily rely on hand-crafted acoustic features, such as Mel-frequency cepstral coefficients (MFCC) and perceptual linear prediction (PLP). These features are designed by domain experts based on the statistical properties of speech signals and perform well in some scenarios. However, as application scenarios become more complex and the scale of data increases dramatically, the limitations of traditional methods become more evident: hand-crafted features lack the ability to generalize across complex data distributions and fail to effectively capture deep acoustic patterns.

In recent years, the advent of deep learning has brought revolutionary changes to the SV field. End-to-end methods based on deep neural networks (DNNs) \cite{Chung2018VoxCeleb2, Chung2020MetricLearning}
automatically extract features and perform classification, eliminating the process of manual feature engineering and significantly improving model performance. However, these methods heavily rely on large-scale, high-quality labeled data, which has become a major bottleneck\cite{stafylakis2019self}. The annotation of speech data is not only costly and time-consuming but is also restricted by data privacy and ethical issues, further limiting the availability of large-scale labeled data\cite{stafylakis2019self, nagrani2020disentangled, chung2020seeing}
.

To address the dependency on labeled data, self-supervised learning (SSL) has gradually become a research focus in the speech recognition field\cite{Cai2021IterativeFramework, Tao2022LossGatedLearning, Sang2022SiameseNetwork}
. SSL generates pseudo-labels from unlabeled data to achieve large-scale feature learning. In these methods, clustering algorithms play a key role in generating pseudo-labels and have shown significant advantages in handling large-scale speech data.

The state-of-the-art self-supervised speaker verification systems are generally divided into two stages. The first stage uses self-supervised learning to extract representations from speech data. The second stage applies clustering algorithms to generate pseudo-labels to further optimize the model. During the representation learning phase, algorithms like DINO\cite{chen2023self}
 (Distilled Instance-level Contrastive Learning) have become mainstream. DINO, as a self-distillation learning framework, aligns the outputs of the teacher and student models to extract stable representations of the same speech segment, effectively reducing the interference of acoustic channel features in representation learning. However, due to the lack of explicit speaker identity information, the representation learning phase still faces some limitations in practical applications.

The second stage utilizes clustering algorithms to perform unsupervised classification of the features generated in the first stage and generates pseudo-labels for subsequent supervised training. While traditional clustering algorithms such as K-Means and agglomerative hierarchical clustering (AHC) perform well on small-scale datasets, they face challenges such as high computational complexity and poor scalability when applied to large-scale data. Moreover, pseudo-labels generated through clustering often contain noisy labels, which can degrade model performance.

To address these issues, graph-based clustering methods have received increasing attention in recent years. In particular, the introduction of graph convolutional networks \cite{Chiang2019ClusterGCN}
has provided new possibilities for clustering tasks on large-scale data. GCN\cite{kinnunen2010overview, lee2011joint, das2018investigating, zhou2021language} constructs similarity graphs between samples, utilizing the internal structure of the data, which not only captures complex relationships between data points but also improves the quality of pseudo-labels. Combining GCN\cite{Schlichtkrull2018GCN}
 with clustering algorithms results in higher precision in pseudo-label generation and significantly enhances the performance of self-supervised speaker verification systems when processing large-scale data.

This study explores the application of the DINO \cite{chen2023self}
framework and graph-based clustering methods based on GCN in speaker verification tasks, with a focus on optimizing the pseudo-label generation process to improve model adaptability to large-scale unlabeled data and evaluating the impact of pseudo-label quality on model performance. Experimental results show that the combination of self-supervised learning and graph-based strategies provides an efficient and scalable solution for speaker verification tasks, while also offering innovative approaches for unsupervised learning in other speech processing domains.

\section{Related Work}
This section systematically reviews and discusses related work in the following two aspects: the application of the DINO \cite{Caron2021SelfSupervisedVT}
self-supervised learning system and graph convolutional network (GCN)-based clustering algorithms.

a) \textbf{Application of DINO (Distilled Instance-level Contrastive Learning) in Speaker Verification}: DINO\cite{Geiping2020CookbookSelfSupervised, He2018TwofoldSiameseNetwork}
is a self-supervised learning method originally designed for image classification tasks in computer vision. Its core innovation lies in discarding negative samples and focusing on feature learning through self-distillation. In speaker verification tasks, DINO \cite{chen2023self}employs a multi-cropping strategy, randomly sampling both short-term and long-term segments from speech clips. Long-term segments capture more stable speaker embeddings, while short-term segments provide finer-grained details.

The DINO framework consists of two branches: a student network and a teacher network. The student network processes all segments, while the teacher network processes only the long-term segments, establishing a "local-to-global" correspondence between short-term and long-term segments. Both networks share the same architecture, but the parameter update methods differ. The student network is updated via gradient descent, while the teacher network is updated using an exponential moving average (EMA) of the student network's parameters. This design effectively mitigates gradient fluctuations and promotes stable feature learning.

Moreover, DINO introduces cosine consistency loss during training, which maximizes the cosine similarity between embeddings of the same speaker, enhancing the discriminative power and stability of the embeddings. By discarding negative samples and combining self-distillation, DINO achieves effective learning for speaker verification tasks, providing a new paradigm for self-supervised learning in speech processing.

b) \textbf{Graph Convolutional Network (GCN)-based Clustering Algorithms}: Clustering is one of the classic unsupervised learning tasks, widely applied in data analysis, classification, and feature learning. Traditional clustering algorithms, such as K-Means, spectral clustering, hierarchical clustering, and DBSCAN, provide a solid theoretical foundation for clustering analysis. However, these methods typically rely on simple assumptions, such as spherical data distributions or density consistency. As a result, when faced with complex, high-dimensional, and unevenly distributed real-world data, their performance is often limited, and they may fail to capture the underlying structure of the data adequately.

In recent years, with the development of deep learning technologies, graph-based clustering algorithms have become a research hotspot. The introduction of Graph Convolutional Networks (GCN) in clustering tasks offers a new approach to solving clustering problems in high-dimensional, complex data. GCN constructs a similarity graph between samples, enabling information propagation across sample nodes and fully utilizing the structural information of the data. Compared to traditional methods, GCN can capture nonlinear relationships between samples, significantly improving clustering accuracy and robustness.

Combining GCN with clustering algorithms not only optimizes clustering performance through high-quality graph representations but also enhances the pseudo-label generation process, providing more reliable supervision signals for self-supervised learning. This integration shows great potential in large-scale speech data processing and feature learning tasks.

\begin{figure*}[htbp]
\centering
\includegraphics[width=\textwidth]{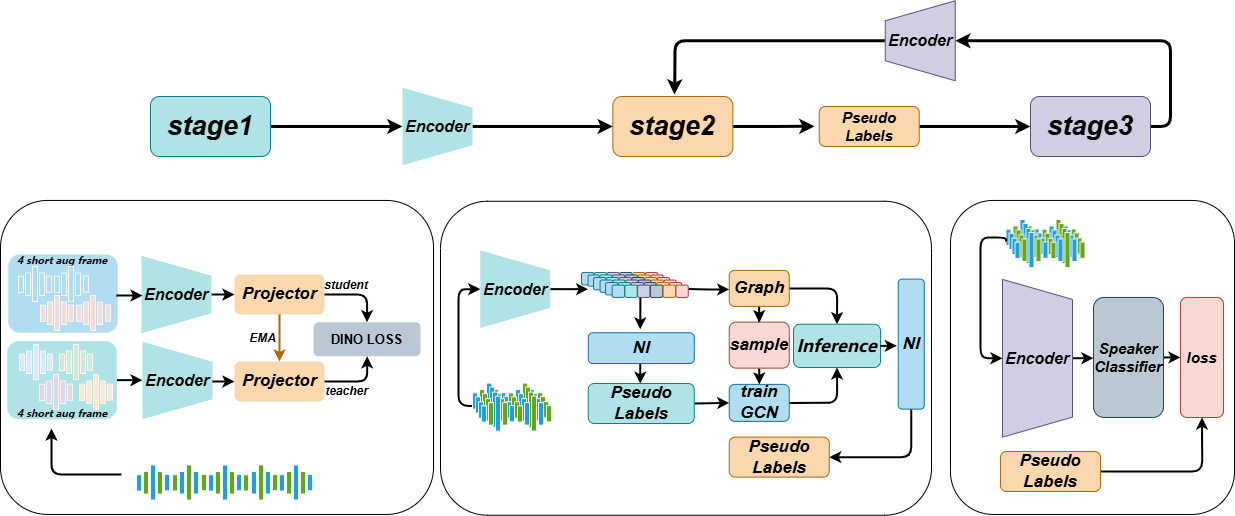}
\caption{Unsupervised Speaker Verification System Architecture
 This figure illustrates the overall architecture of the unsupervised speaker verification system, consisting of three main stages: In the first stage, the initial encoder is extracted through the DINO self-distillation framework to generate the initial embeddings. In the second stage, a clustering method is applied to generate pseudo-labels, and GCN is used for training to optimize the graph structure and identify precise edge connections. In the third stage, the quality of pseudo-labels is optimized, improving the training performance of the backend speaker verification model, and the encoder is iteratively updated to enhance system performance.
}
\label{fig:twocolumn}
\end{figure*}

\section{Method}
Fig.1 illustrates the overall structure of the proposed unsupervised speaker verification system, which consists of three main stages.

\textbf{Stage 1}: In this stage, we employ the DINO\cite{chen2023self}
 self-distillation framework as the initial training tool, aiming to uncover latent feature representations from large-scale unlabeled data. The DINO framework efficiently extracts embedding representations of speech samples through a contrastive learning approach, without relying on any label information, thereby providing high-quality feature representations for subsequent model training. Specifically, we first train an initial encoder using the DINO framework, which generates preliminary embedding representations of the data. These preliminary embeddings provide an important feature foundation for subsequent KNN-based  \cite{cover1967nearest}
similarity graph construction and further training. Therefore, this stage lays a solid foundation for the entire system's feature extraction, ensuring the efficiency and accuracy of the subsequent stages.

\textbf{Stage 2}: In this stage, we construct a similarity graph based on the initial embeddings generated by DINO and apply the proposed clustering method to generate initial pseudo-labels. Given that the clustering method is highly robust, we can select a batch of higher-quality pseudo-labels from the initial ones for subsequent GCN training. The GCN training process adopts a class-based sampling strategy, using these pseudo-labels to sample the entire graph and generate multiple subgraphs. This strategy allows for the generation of richer subgraphs while preserving key structural information from the overall graph, even with a limited amount of data. These subgraphs and their corresponding node embeddings are then fed into a Graph Convolutional Network (GCN) for training. The GCN, with its excellent aggregation ability, effectively captures the relationships between nodes and performs edge classification using a cross-entropy loss function. In this way, the GCN model can precisely identify and optimize the edge connections in the graph structure during the training process. During the inference phase, the entire graph is input into the trained GCN model, which evaluates the edge connections, simplifying the graph structure and removing erroneous connections. We then apply the clustering method based on the similarity graph once again to obtain a batch of high-quality pseudo-labels for the subsequent speaker verification model training.

\textbf{Stage 3}: In this stage, we train a speaker verification model based on the high-quality pseudo-labels obtained in Stage 2. Unlike Stage 1, where the DINO framework is used for training, Stage 3’s model training relies on high-quality pseudo-labels, significantly improving the model's performance. This approach enables the encoder to more efficiently capture speaker features, resulting in more accurate embedding representations. Consequently, we use the newly trained encoder from Stage 3 to extract higher-quality speaker embeddings and use them as input to repeat the process from Stage 2, further optimizing the quality of the pseudo-labels.

\subsection{DINO}
DINO is a self-distillation framework designed to learn high-quality audio features in the absence of labeled data. The framework consists of a student network and a teacher network. Given an audio input \( x \), data augmentation generates global clips \( x_g \) and local clips \( x_l \). The teacher network only receives the global clip \( x_g \), while the student network receives all clips \( x \). The outputs of both networks are normalized via softmax to obtain the probability distributions \( P_s(x) \) for the student network and \( P_t(x) \) for the teacher network.

To learn the local-to-global relationships, DINO minimizes the cross-entropy loss \( L_{DINO} \) to align the output of the student network with that of the teacher network. Specifically, the goal is to minimize the cross-entropy between the outputs of the student network and the teacher network:

\begin{equation}
L_{DINO} = \sum_{x \in x_g} \sum_{x' \in X \setminus \{x\}} \text{CE}(P_s(x') \parallel P_t(x))
\end{equation}

Here, \( P_s(x') \) and \( P_t(x) \) represent the probability distributions from the student and teacher networks, respectively, and \( \text{CE}(P_s(x') \parallel P_t(x)) \) denotes the cross-entropy between the output \( P_s(x') \) of the student network and the output \( P_t(x) \) of the teacher network. The term \( X \setminus \{x\} \) refers to the set of all samples except the current clip \( x \).

The cross-entropy formula is given by:
\begin{equation}
\text{CE}(P_s(x') \parallel P_t(x)) = - \sum_{k=1}^{K} P_t(x_k) \log P_s(x')
\end{equation}

To ensure the stability of the network's output, DINO uses a softmax function with a temperature parameter \( \tau \) for normalization. The temperature parameter \( \tau \) controls the smoothness of the output distribution, with larger temperatures leading to a smoother distribution, thereby enhancing the robustness of the model. The temperature softmax formula is:

\begin{equation}
P(x) = \frac{\exp\left(\frac{g_\theta(x)}{\tau}\right)}{\sum_{k=1}^{K} \exp\left(\frac{g_\theta(x_k)}{\tau}\right)}
\end{equation}

Here, \( g_\theta(x) \) represents the feature representation obtained from the network, and \( \tau \) controls the sharpness of the output. The temperature softmax helps balance the sharpness and smoothness of the output by adjusting \( \tau \), enabling the model to stably learn more expressive features.

This mechanism plays a crucial role in ensuring that the student network learns features consistent with the output of the teacher network, even when no labeled data is available.

\subsection{GCN Network Architecture }
In this paper, we design a framework based on Graph Convolutional Networks (GCN) to learn the edge connection relationships between nodes and optimize the similarity matrix through the binary classification of edges. The GCN network consists of multiple convolutional layers, each of which updates node representations by aggregating information from neighboring nodes, progressively capturing local structures and dependencies in the graph. The input to the GCN includes the adjacency matrix \( A \) and the node feature matrix \( X \). The adjacency matrix \( A \) describes the connection relationships between nodes in the graph, while the node feature matrix \( X \) contains the original embedding representations of the nodes. Through graph convolution operations, the node representations at each layer are updated based on information from neighboring nodes.

The update rule for the node representations at the \( l \)-th layer is as follows:

\begin{equation}
H^{(l+1)} = \sigma\left( \hat{A}H^{(l)}W^{(l)} + XW_{\text{input}}^{(l)} \right)
\end{equation}

where \( H^{(l)} \in \mathbb{R}^{N \times D_l} \) is the node feature matrix at the \( l \)-th layer, representing the node representations at the current layer; \( X \in \mathbb{R}^{N \times D_{\text{input}}} \) is the input node feature matrix, containing the initial embedding representations of each node; \( W^{(l)} \in \mathbb{R}^{D_l \times D_{l+1}} \) is the learnable weight matrix at the \( l \)-th layer, responsible for updating the node representations; \( W_{\text{input}}^{(l)} \in \mathbb{R}^{D_{\text{input}} \times D_l} \) is the learnable weight matrix for the input feature matrix; and \( \hat{A} = D^{-1/2}AD^{-1/2} \) is the normalized adjacency matrix, where \( A \) is the original adjacency matrix and \( D \) is the degree matrix.

In this formula, \( \hat{A}H^{(l)}W^{(l)} \) represents the propagation of information from neighboring nodes through the graph structure, while \( XW_{\text{input}}^{(l)} \) applies a linear transformation to the input features. The node representations are then updated through the activation function \( \sigma(\cdot) \) (typically ReLU). The node representations at the 0-th layer are initialized as the input feature matrix, i.e., \( H^{(0)} = X \).

We further define the characteristics of the graph convolution operation, including the information propagation between nodes. At the \( l \)-th layer, the update rule for node \( i \) can be expressed as:

\begin{equation}
h_i^{(l+1)} = \sigma\left( \sum_{j \in \mathcal{N}(i)} \frac{1}{\sqrt{d_i d_j}} W^{(l)} h_j^{(l)} + W_{\text{input}}^{(l)} x_i \right)
\end{equation}

where \( h_i^{(l)} \) represents the feature of node \( i \) at the \( l \)-th layer, \( \mathcal{N}(i) \) is the set of neighboring nodes of node \( i \), \( d_i \) and \( d_j \) are the degrees of nodes \( i \) and \( j \), \( W^{(l)} \) is the learnable weight at the \( l \)-th layer, and \( x_i \) is the input feature of node \( i \).

By stacking multiple GCN layers, node representations progressively aggregate information from neighbors, capturing local structures and dependencies in the graph. This process allows the model to learn richer node representations based on the graph structure.

Finally, the learned node representations can be used for the binary classification of edges. For edge classification, we use the following formula to compute the probability of edge existence:

\begin{equation}
p_{ij} = \sigma\left( \mathbf{w}^T [h_i \oplus h_j] \right)
\end{equation}

where \( p_{ij} \) is the probability of the existence of an edge between nodes \( i \) and \( j \), \( h_i \) and \( h_j \) are the representations of nodes \( i \) and \( j \), \( \oplus \) denotes the concatenation of node representations, and \( \mathbf{w} \) is the learnable weight vector. We train the network using a cross-entropy loss function to optimize the fusion of node features and graph structural information.

Through this process, graph convolution not only effectively captures the local structural features between nodes but also automatically extracts useful representations from the graph structure, thus improving the performance of downstream tasks.
\subsection{Training }
The GCN we designed employs supervised training; however, due to the lack of labeled data, the generation of high-quality pseudo-labels is crucial for initializing the GCN training process. To address this, we propose a similarity-based clustering method, which demonstrates strong capabilities. Specifically, we first use the encoder of the pre-trained DINO model to extract speaker embeddings for all the data and construct an N×N similarity matrix using the KNN\cite{cover1967nearest}.
 algorithm (where N represents the number of samples). Next, we compute the similarity of all connected nodes in the similarity matrix and remove low-similarity connections based on a predefined threshold. Then, based on the processed similarity matrix, we apply the proposed clustering method to generate initial pseudo-labels.
\begin{figure}[!h]
    \centering
    \includegraphics[width=1\linewidth]{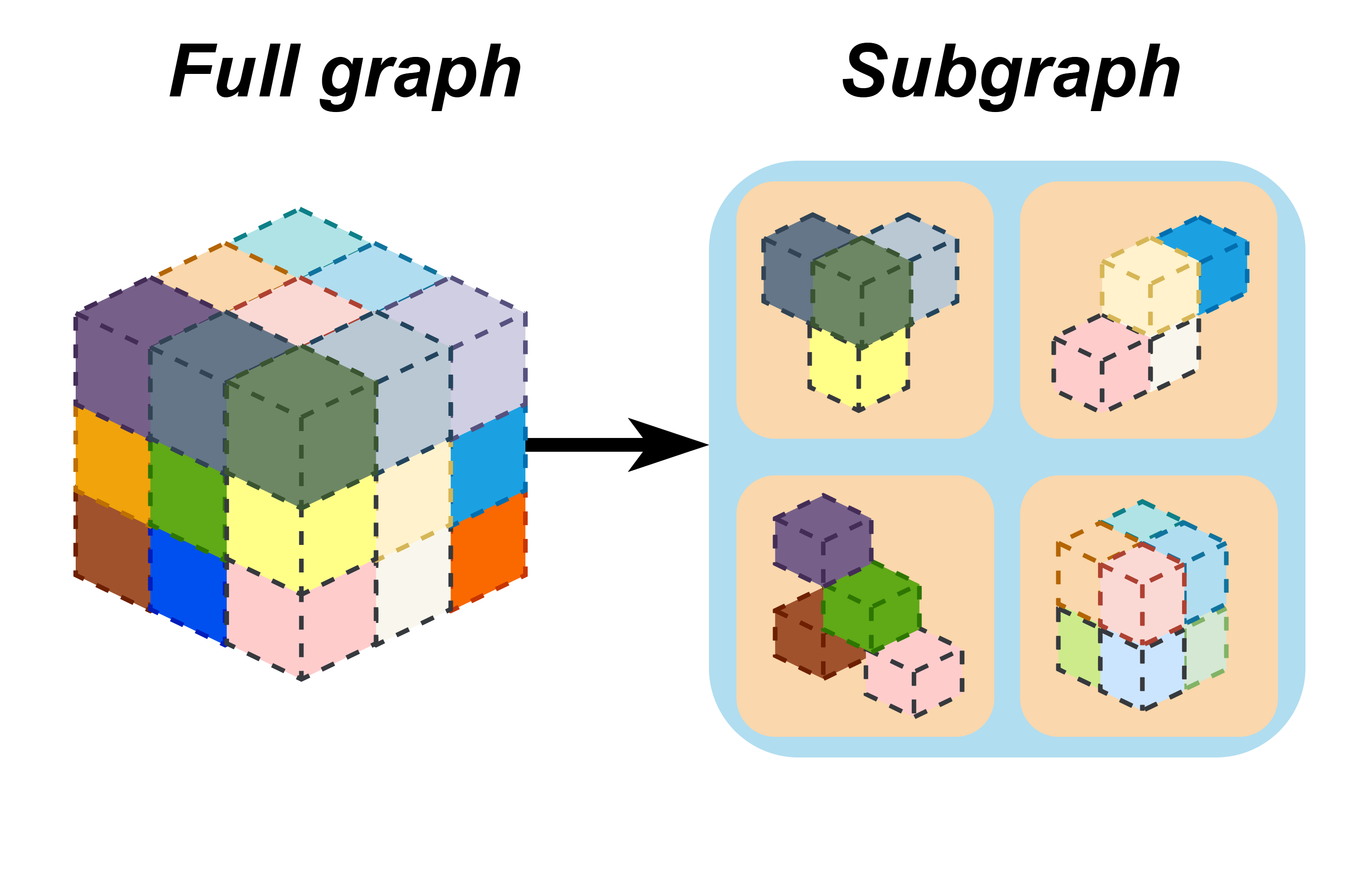}
    \caption{Subgraph sampling illustration: Each differently colored cube in the figure represents the sample space of a specific class. Based on the center coordinates of each class, we select the n nearest classes to the current class center for sampling and construct the corresponding subgraphs for subsequent training and analysis.
}
    \label{fig:enter-label}
\end{figure}

Since not all data is needed for GCN training, we set a selection threshold to retain only those classes that meet the threshold condition (approximately one-quarter of the total data), and the data corresponding to these classes will be used for GCN training.

To meet the GCN's requirement for large-scale graphs, we adopt a full-graph sampling strategy. As random node sampling may fail to ensure necessary cross-class similarity connections, we design a class-based sampling method with the following steps:

\begin{enumerate}
    \item Randomly select n1 classes;
    \item From each selected class, randomly choose n2 samples. Within the selected k samples, KNN \cite{cover1967nearest}is used to build a proximity matrix and generate a high-quality subgraph, ensuring the subgraph structure meets the input requirements for the GCN.
\end{enumerate}
\begin{figure}[!h]
    \centering
    \includegraphics[width=1\linewidth]{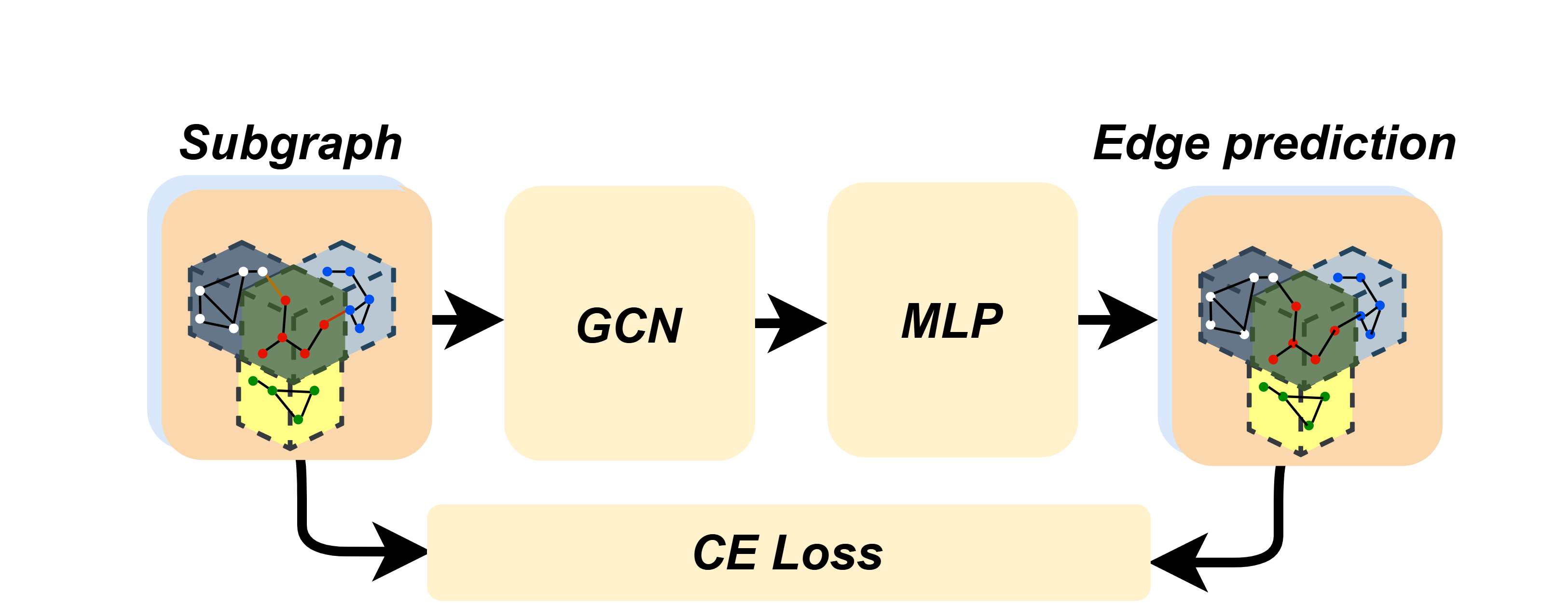}
    \caption{GCN training process: The sampled subgraph is input into the GCN, which aggregates neighboring information layer by layer through graph convolution, updating node representations and capturing local structures. An MLP predicts the edges between node pairs, and the training is optimized using a cross-entropy loss function to improve edge classification accuracy.}
    \label{fig:enter-label}
\end{figure}
We then feed the sampled subgraph into the GCN, which aggregates information from neighboring nodes layer by layer through graph convolution operations. This process updates node representations and progressively captures local structures and dependencies between nodes. The node representations at each layer are updated based on information from neighboring nodes until they contain rich structural information. Subsequently, an MLP \cite{Jain1996ANN}
processes these node representations to predict whether an edge exists between each pair of nodes. The training process is optimized using a cross-entropy loss function to improve the accuracy of edge classification.

\subsection{Inference }

In the clustering process, we first construct a complete similarity connection graph and input it into a trained Graph Convolutional Network (GCN) to predict the connections of each edge. Although the labels in the training data may not be entirely accurate and some misclassifications may occur during inference, the number of positive samples in the graph is far higher than that of negative samples due to the high similarity between different samples of the same speaker. After GCN inference, the number of incorrect connections is significantly reduced, thereby improving the accuracy of edge connections. Although this approach may lead to the loss of a small number of correct connections, the overall performance is significantly optimized.
\begin{figure}[!h]
    \centering
    \includegraphics[width=0.75\linewidth]{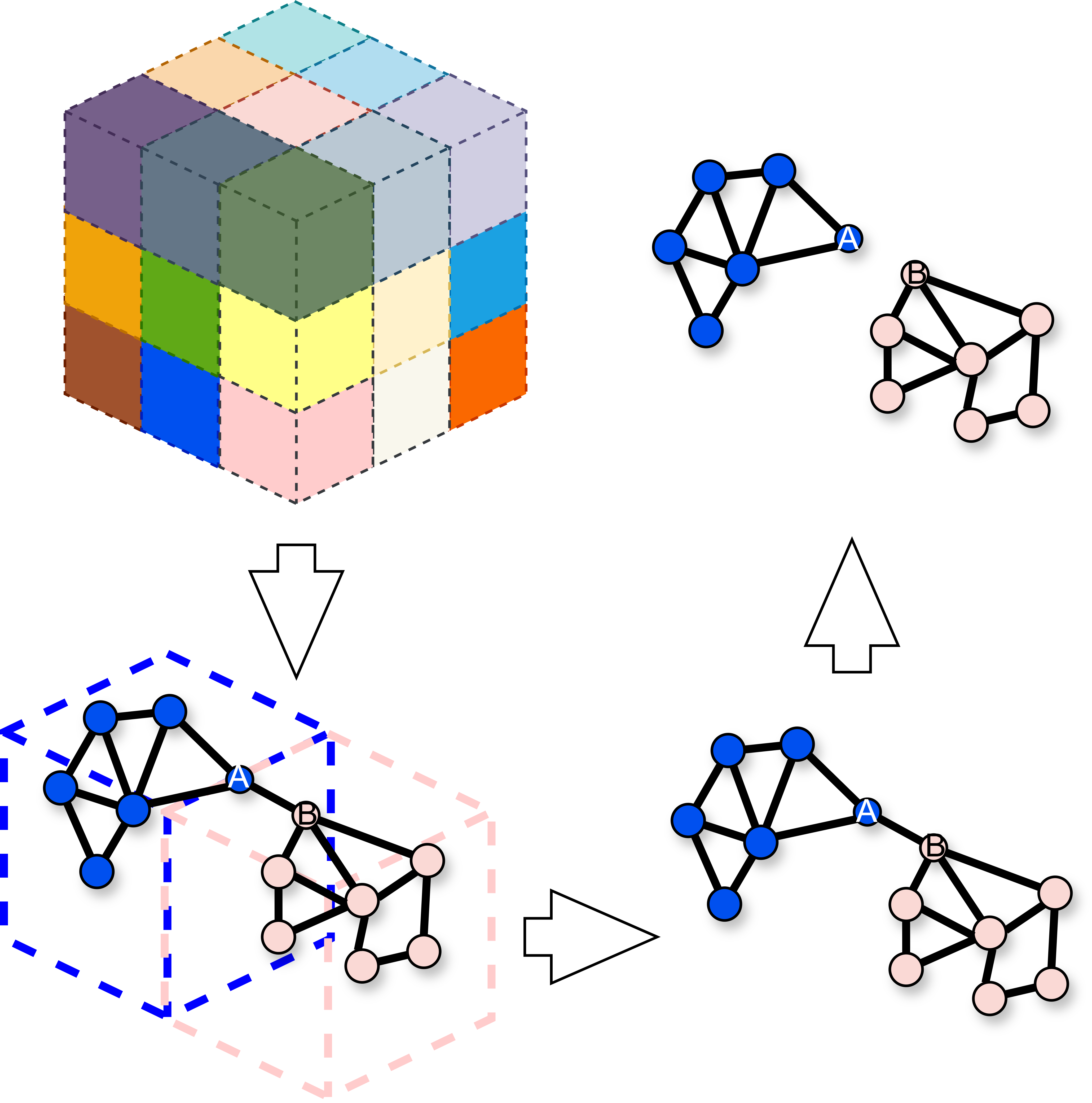}
    \caption{This figure illustrates the specific details of the inference process, where the connections between samples represent the distance (similarity) between each pair. Although two samples from different categories may sometimes have high similarity, the lack of common neighbors between them reduces the weight of their similarity, thereby affecting the overall similarity calculation and ultimately achieving the distinction between different categories.
}
    \label{fig:enter-label}
\end{figure}

However, when calculating the similarity between nodes in the graph, we found that the similarity between nodes is not only influenced by the features of the nodes themselves but also by their neighbor structures. To improve the accuracy of similarity calculation more effectively, we propose a simplified weighted similarity method, which focuses on the influence of common similar neighbors between nodes. This method not only improves the accuracy of node self-similarity calculations but also fully utilizes the adjacency relationships between nodes, resulting in better performance in classification or clustering tasks.

We believe that the similarity between nodes depends not only on the nodes themselves but also on their common neighbors. In particular, when nodes \(i\) and \(j\) share many common neighbors, the similarity of these common neighbors plays a stronger guiding role in determining the relationship between the two nodes. Therefore, to better reflect the influence of common neighbors on node similarity, we introduce a weighting term. The specific calculation formula is as follows:

\begin{equation}
C(i, j) = \textstyle\sum_{k \in N_{\text{common}}(i,j)} (\alpha_i S(i, k) + \alpha_j S(j, k))
\end{equation}
where
\[
\alpha_i = \frac{|N_{common}(i,j)|}{|N(i)|}, \quad \alpha_j = \frac{|N_{common}(i,j)|}{|N(j)|}
\]
are the weighting coefficients for nodes \(i\) and \(j\). \(|N(i)|\) and \(|N(j)|\) are the total number of neighbors of nodes \(i\) and \(j\), respectively, while \(|N_{common}(i, j)|\) is the number of common similar neighbors between nodes \(i\) and \(j\). \(S(i, k)\) and \(S(j, k)\) are the similarities between node \(i\) and node \(k\), and between node \(j\) and node \(k\), respectively.

Through this weighted similarity formula, we can more precisely capture the influence of common similar neighbors on the relationship between nodes, thus effectively improving the accuracy of classification and clustering.

The advantages of this method are as follows:
\begin{itemize}
    \item Improved similarity calculation accuracy: By emphasizing the influence of common similar neighbors, the accuracy of similarity calculation between nodes is significantly improved, especially in cases where the similarity distribution is uneven.
    \item Flexible adjustment based on neighbor structure: The inclusion of the weighting term for the number of neighbors allows for flexible adjustments to the calculation based on the neighbor structure of the nodes, avoiding potential biases that may arise from relying solely on the similarity between nodes.
\end{itemize}

This method allows for better handling of node relationships in complex network structures, leading to improved clustering and classification accuracy.

\section{EXPERIMENTS }
\subsection{Experimental setup}
• \textbf{Dataset}: This experiment uses the VoxCeleb2~\cite{Chung2018VoxCeleb2} dataset, which contains 1,092,009 speech samples from 5,994 speakers. No accurate label information was used during the entire training process. Additionally, data augmentation was performed using MUSAN~\cite{david2015musan} noise and reverberation corpora from the Room Impulse Response and Noise~\cite{ko2017data} (RIRs) database.

• \textbf{DINO}: The training process follows the settings recommended in the literature, using the ECAPA-TDNN architecture as the encoder to generate 512-dimensional embedding vectors.

• \textbf{GCN}: During training, 300,000 data points were used. The generated pseudo-labels consist of 5,998 classes (while the real labels have 5,964 classes), with the Normalized Mutual Information (NMI) between pseudo-labels and real labels reaching 96.64.

• \textbf{Similarity Graph Construction}: In the experiment, we used the faiss-gpu library to construct a KNN \cite{cover1967nearest}similarity graph with K set to 50, meaning each sample is connected to the 50 nearest samples based on Euclidean distance.

• \textbf{Backend}: The backend uses the ECAPA-TDNN\cite{desplanques2020ecapa}
 (512-dimensional) architecture. The margin in the AAM is set to 0.2, and the scaling factor is set to 32. The optimization process uses the Adam optimizer, combined with a cyclic learning rate (CLR) strategy, with the learning rate range set from 1e-3 to 1e-6. At the beginning of training, a simple threshold is set to remove pseudo-label samples with fewer than 20 samples in each class, ensuring that around 1,040,000 valid samples are used for training in each iteration.
\subsection{Results and discussion}
\begin{table*}[!ht]
    \caption{\textbf{Comparison with Results of Other Methods:}}
    \centering
    \begin{tabular*}{\textwidth}{@{\extracolsep{\fill}}llccccccc}
        \toprule
        \textbf{Methods} & \textbf{Model} & \textbf{Iteration} & \textbf{Cluster} & \textbf{Clusters} & \textbf{NMI} & \textbf{Vox-O} & \textbf{Vox-E} & \textbf{Vox-H} \\
        \midrule
        DINO & ECAPA-S & - & - & - & - & 3.16 & - & - \\ 
        ID\cite{thienpondt2020idlab} & ECAPA-L & 7 & AHC & 7500 & - & 2.10 & - & - \\ 
        JHU\cite{cho2021jhu} & Res2Net50 & 4 & AHC & 7500 & - & 1.89 & - & - \\ 
        DKU\cite{cai2021dku} & ResNet & 4 & K-M & 6000 & 95.20 & 1.81 & - & - \\ 
        SNU\cite{mun2021snu} & ECAPA-L & 4 & AHC & 7500 & - & 1.66 & - & - \\ 
        LGL\cite{tao2021self} & ECAPA-L & 5 & K-M & 6000 & 93.32 & 1.66 & 2.18 & 3.76 \\ 
        LGL & ECAPA-S & - & K-M & 7500 & - & 2.02 & - & - \\ 
        DLG-LC\cite{bing2022self} & ECAPA-S & 3 & K-M & 7500 & - & 1.67 & - & - \\ 
        MiniBatch P C\cite{Wang2024SelfSupervisedSV} & ECAPA-L & - & AHC & 6000 & - & 1.70 & - & - \\ 
        OURS & ECAPA-S & 3 & OURS & - & 97.04 & 1.57 & 2.01 & 3.46 \\
        \bottomrule
    \end{tabular*}
    \label{tab:combined}
\end{table*}
In this section, we rigorously demonstrate the superiority of our proposed method by comparing it with several typical unsupervised learning methods. Traditional clustering methods often require pre-setting the number of clusters, which is typically based on experience or tuning and may lead to biased or unstable clustering results. In contrast, our method utilizes self-supervised learning for high-quality feature extraction and optimizes the clustering strategy through a Graph Convolutional Network (GCN), allowing it to automatically determine the optimal number of clusters. This strategy effectively avoids the uncertainty and bias that may arise from manually setting the number of clusters.
Our experimental results show that the OURS method performs excellently across multiple validation sets. For example, on the Vox-O validation set, our method achieves an EER of 1.57, significantly lower than other methods. Notably, we did not introduce additional optimization techniques, such as pseudo-label correction, in the backend model, and our method still significantly outperforms others by solely relying on the improved clustering results. This indicates that our method demonstrates strong capability in distinguishing speaker differences and achieves clustering performance that clearly surpasses traditional methods.

Further analysis shows that our method optimizes pseudo-label generation via GCN, significantly enhancing the clustering structure and resulting in a pseudo-label NMI of 97.04, which is much higher than other methods, such as LGL (93.32) and DKU (95.20). This high-quality pseudo-label not only improves clustering performance but also provides solid support for subsequent training.
\begin{table}[!ht]
\captionsetup{justification=raggedright, singlelinecheck=false}
\centering
\renewcommand{\arraystretch}{1.2}
\caption{\textbf{Ablation Results of Similarity Calculation .} Method 1: $S(i, k) \cdot S(j, k)$; Method 2: $S(i, k) + S(j, k)$; Method 3: $\alpha_i S(i, k) + \alpha_j S(j, k)$; Method 4: GCN-enhanced weighted similarity.}
\begin{tabular}{>{\centering\arraybackslash}p{0.5\linewidth}l>{\centering\arraybackslash}p{0.3\linewidth}}
    \toprule
    \textbf{Method} & \textbf{K} & \textbf{NMI} \\
    \midrule
    K-Means (6000 clusters) & – & 93.6 \\
    Method 1 & 50 & 93.7 \\
    Method 2 & 50 & 93.9 \\
    Method 3 & 50 & 95.4 \\
    Method 4 & 20 & 92.9 \\
    Method 4 & 50 & 96.6 \\
    Method 4 & 80 & 96.7 \\
    Method 4 & 100 & 90.2 \\
    \bottomrule
\end{tabular}
\end{table}
\FloatBarrier

In summary, our method demonstrates that high-quality clustering results can be achieved without manually specifying the number of clusters. By relying on automated clustering strategies and high-quality feature extraction, it achieves efficient and accurate clustering results. Across multiple validation sets, our method outperforms other methods in terms of EER and NMI, fully validating its effectiveness and superiority in unsupervised speaker verification tasks.

 \subsection{Clustering Results}
In this experiment, we investigated the impact of different similarity calculation methods on clustering performance. First, we used the traditional K-Means clustering algorithm with 6000 clusters as a baseline, achieving an NMI of 93.6. Next, we introduced three methods based on common nodes to improve clustering results. Method 1 calculates similarity as $S(i,k) \cdot S(j,k)$, with an NMI of 93.7, which is similar to the K-Means result. Method 2 calculates similarity as $S(i,k) + S(j,k)$, resulting in an NMI of 93.9, with no significant improvement. Method 3 introduced weighted similarity $\alpha_i S(i,k) + \alpha_j S(j,k)$, which increased the NMI to 95.4, indicating that accounting for the weight of common nodes significantly improved the accuracy of similarity calculation. In Method 4, we combined Graph Convolutional Networks  with the weighted similarity method, further improving clustering performance. The results with different values of common nodes K show that when K is 20, the NMI is 92.9, and when K is 50, the NMI is 96.6, which is similar to the result when K is 80. However, when K is increased to 100, the NMI drops to 90.2, indicating a degradation in clustering performance. We believe that too many common nodes may lead to incorrect connections, affecting the clustering accuracy. Therefore, we selected K=50 as the optimal value while avoiding the higher computational cost of K=80. traditional K-Means clustering. 
\section{Conclusion}
The improved clustering framework based on similarity connection graphs and Graph Convolutional Networks proposed in this paper significantly enhances the performance of self-supervised speaker verification systems by optimizing the clustering process and improving the accuracy of pseudo-labels. Traditional clustering methods tend to be affected by noisy data and inaccurate labels when handling complex speaker verification tasks, leading to performance degradation. In contrast, this method effectively captures the latent similarity relationships between speakers by leveraging the powerful expressive capabilities of Graph Convolutional Networks, thereby better utilizing the intrinsic structure of the data during the clustering process and improving the quality of pseudo-labels. Experimental results show that this method outperforms traditional clustering methods in several evaluation metrics.
\section{Limitations}
The current performance of the system is limited by the quality of the embeddings. We are actively improving the performance of DINO to achieve optimal results with the fewest possible training iterations.
\newpage
\bibliographystyle{IEEEtran}
\bibliography{references}

\begin{thebibliography}{10}
\providecommand{\url}[1]{#1}
\csname url@samestyle\endcsname
\providecommand{\newblock}{\relax}
\providecommand{\bibinfo}[2]{#2}
\providecommand{\BIBentrySTDinterwordspacing}{\spaceskip=0pt\relax}
\providecommand{\BIBentryALTinterwordstretchfactor}{4}
\providecommand{\BIBentryALTinterwordspacing}{\spaceskip=\fontdimen2\font plus
\BIBentryALTinterwordstretchfactor\fontdimen3\font minus \fontdimen4\font\relax}
\providecommand{\BIBforeignlanguage}[2]{{%
\expandafter\ifx\csname l@#1\endcsname\relax
\typeout{** WARNING: IEEEtran.bst: No hyphenation pattern has been}%
\typeout{** loaded for the language `#1'. Using the pattern for}%
\typeout{** the default language instead.}%
\else
\language=\csname l@#1\endcsname
\fi
#2}}
\providecommand{\BIBdecl}{\relax}
\BIBdecl

\bibitem{Chung2018VoxCeleb2}
J.~S. Chung, A.~Nagrani, and A.~Zisserman, ``Voxceleb2: Deep speaker recognition,'' in \emph{Interspeech}, 2018, pp. 1086--1090.

\bibitem{Chung2020MetricLearning}
J.~S. Chung, J.~Huh, S.~Mun, M.~Lee, H.-S. Heo, S.~Choe, C.~Ham, S.~Jung, B.-J. Lee, and I.~Han, ``In defence of metric learning for speaker recognition,'' in \emph{Interspeech}, 2020.

\bibitem{stafylakis2019self}
T.~Stafylakis, J.~Rohdin, O.~Plchot, P.~Mizera, and L.~Burget, ``Self-supervised speaker embeddings,'' in \emph{Interspeech}, 2019, pp. 2863--2867.

\bibitem{nagrani2020disentangled}
A.~Nagrani, J.~S. Chung, S.~Albanie, and A.~Zisserman, ``Disentangled speech embeddings using cross-modal self-supervision,'' in \emph{ICASSP}, 2020, pp. 6829--6833.

\bibitem{chung2020seeing}
S.-W. Chung, H.-G. Kang, and J.~S. Chung, ``Seeing voices and hearing voices: Learning discriminative embeddings using cross-modal self-supervision,'' in \emph{Interspeech}, 2020, pp. 3486--3490.

\bibitem{Cai2021IterativeFramework}
D.~Cai, W.~Wang, and M.~Li, ``An iterative framework for self-supervised deep speaker representation learning,'' in \emph{ICASSP}, 2021, pp. 6728--6732.

\bibitem{Tao2022LossGatedLearning}
R.~Tao, K.~A. Lee, R.~K. Das, V.~Hautamaki, and H.~Li, ``Self-supervised speaker recognition with loss-gated learning,'' in \emph{ICASSP}, 2022, pp. 6142--6146.

\bibitem{Sang2022SiameseNetwork}
M.~Sang, H.~Li, F.~Liu, A.~O. Arnold, and L.~Wan, ``Self-supervised speaker verification with simple siamese network and self-supervised regularization,'' in \emph{ICASSP}, 2022, pp. 6127--6131.

\bibitem{chen2023self}
Y.~Chen, S.~Zheng, H.~Wang, L.~Cheng, and Q.~Chen, ``Pushing the limits of self-supervised speaker verification using regularized distillation framework,'' in \emph{ICASSP}, 2023.

\bibitem{Chiang2019ClusterGCN}
W.-L. Chiang, X.~Liu, S.~Si, Y.~Li, S.~Bengio, and C.-J. Hsieh, ``Cluster-gcn: An efficient algorithm for training deep and large graph convolutional networks,'' in \emph{KDD}, 2019.

\bibitem{kinnunen2010overview}
T.~Kinnunen and H.~Li, ``An overview of text-independent speaker recognition: From features to supervectors,'' \emph{Speech Communication}, vol.~52, pp. 12--40, 2010.

\bibitem{lee2011joint}
K.~A. Lee, A.~Larcher, H.~Thai, B.~Ma, and H.~Li, ``Joint application of speech and speaker recognition for automation and security in smart home,'' in \emph{Interspeech}, 2011, pp. 3317--3318.

\bibitem{das2018investigating}
R.~K. Das and S.~R.~M. Prasanna, ``Investigating text-independent speaker verification from practically realizable system perspective,'' in \emph{APSIPA ASC}, 2018, pp. 1483--1487.

\bibitem{zhou2021language}
Y.~Zhou, X.~Tian, and H.~Li, ``Language agnostic speaker embedding for cross-lingual personalized speech generation,'' \emph{IEEE/ACM Transactions on Audio, Speech, and Language Processing}, vol.~29, pp. 3427--3439, 2021.

\bibitem{Schlichtkrull2018GCN}
M.~Schlichtkrull, T.~N. Kipf, P.~Bloem, R.~V.~D. Berg, I.~Titov, and M.~Welling, ``Modeling relational data with graph convolutional networks,'' in \emph{ESWC}, 2018.

\bibitem{Caron2021SelfSupervisedVT}
M.~Caron, H.~Touvron, I.~Misra, H.~Jegou, J.~Mairal, P.~Bojanowski, and A.~Joulin, ``Emerging properties in self-supervised vision transformers,'' in \emph{ICCV}, 2021, pp. 9657--9667.

\bibitem{Geiping2020CookbookSelfSupervised}
J.~Geiping, Q.~Garrido, P.~Fernandez, A.~Bar, H.~Pirsiavash, Y.~LeCun, and M.~Goldblum, ``A cookbook of self-supervised learning,'' \emph{arXiv preprint arXiv:2003.00168}, 2020.

\bibitem{He2018TwofoldSiameseNetwork}
A.~He, C.~Luo, X.~Tian, and W.~Zeng, ``A twofold siamese network for real-time object tracking,'' in \emph{CVPR}, 2018, pp. 4834--4843.

\bibitem{cover1967nearest}
T.~Cover and P.~Hart, ``Nearest neighbor pattern classification,'' \emph{IEEE Transactions on Information Theory}, vol.~13, no.~1, pp. 21--27, 1967.

\bibitem{Jain1996ANN}
A.~K. Jain, J.~Mao, and K.~M. Mohiuddin, ``Artificial neural networks: A tutorial,'' \emph{Computer}, vol.~29, no.~3, pp. 31--44, 1996.

\bibitem{david2015musan}
S.~David, C.~Guoguo, and P.~Daniel, ``Musan: A music, speech, and noise corpus,'' \emph{arXiv preprint arXiv:1510.08484}, 2015.

\bibitem{ko2017data}
T.~Ko, V.~Peddinti, D.~Povey, M.~L. Seltzer, and S.~Khudanpur, ``A study on data augmentation of reverberant speech for robust speech recognition,'' in \emph{ICASSP}, 2017.

\bibitem{desplanques2020ecapa}
B.~Desplanques, J.~Thienpondt, and K.~Demuynck, ``Ecapa-tdnn: Emphasized channel attention, propagation and aggregation in tdnn based speaker verification,'' in \emph{Interspeech}, 2020.

\bibitem{thienpondt2020idlab}
J.~Thienpondt, B.~Desplanques, and K.~Demuynck, ``The idlab voxceleb speaker recognition challenge 2020 system description,'' \emph{arXiv preprint arXiv:2010.12468}, 2020.

\bibitem{cho2021jhu}
J.~Cho, J.~Villalba, and N.~Dehak, ``The jhu submission to voxsrc-21: Track 3,'' \emph{arXiv preprint arXiv:2109.13425}, 2021.

\bibitem{cai2021dku}
D.~Cai and M.~Li, ``The dku-dukeece system for the self-supervision speaker verification task of the 2021 voxceleb speaker recognition challenge,'' \emph{arXiv preprint arXiv:2109.02853}, 2021.

\bibitem{mun2021snu}
S.~H. Mun, M.~H. Han, and N.~S. Kim, ``Snu-hil system for the voxceleb speaker recognition challenge 2021,'' \emph{VoxSRC}, 2021.

\bibitem{tao2021self}
R.~Tao, K.~A. Lee, R.~K. Das, V.~Hautamaki, and H.~Li, ``Self-supervised speaker recognition with loss-gated learning,'' \emph{arXiv preprint arXiv:2110.03869}, 2021.

\bibitem{bing2022self}
H.~Bing, C.~Zhengyang, and Q.~Yanmin, ``Self-supervised speaker verification using dynamic loss-gate and label correction,'' in \emph{Interspeech}, 2022.

\bibitem{Wang2024SelfSupervisedSV}
J.~Wang, Z.~Fang, and L.~He, ``Self-supervised speaker verification with mini-batch prediction correction,'' \emph{Interspeech}, 2024.

\end{thebibliography}

\end{document}